\documentclass[final,5p,times]{elsarticle}

\usepackage{floatrow}
\usepackage{graphicx}
\usepackage{stfloats}
\usepackage[labelfont=bf,list=true]{subcaption}
\graphicspath{{Images/}}			
\usepackage[font=small,labelfont=bf,labelsep=period]{caption}

\usepackage{lineno}
\usepackage{makecell}
\usepackage{booktabs}
\usepackage{indentfirst}
\usepackage[dvipsnames,svgnames,x11names]{xcolor}

\biboptions{round,numbers,sort&compress}

\usepackage[breaklinks=true]{hyperref}	

\usepackage{natbib}
\setcitestyle{square}

\usepackage{amssymb}
\usepackage{amsmath}
\usepackage{upgreek} 		

\usepackage[nolist,nohyperlinks]{acronym}

\usepackage{booktabs}                  	
\usepackage{tabulary}
\usepackage{array}
\usepackage{multirow}
\usepackage{colortbl}

\usepackage{stfloats}					

\usepackage[nolist]{acronym}

\usepackage[noabbrev,capitalise]{cleveref}	

\begin{document}

\begin{acronym}
\acro{AdEPT}[AdEPT]{Advanced Energetic Pair Telescope}
\acro{CNN}[CNN]{Convolutional Neural Network}
\acro{TPC}[TPC]{Time Projection Chamber}
\acro{GCR}[GCR]{Galactic Cosmic Ray}
\acro{NiN}[NiN]{Network in Network}
\acro{SPENVIS}[SPENVIS]{Space Environment Information System}
\acro{SGD}[SGD]{Stochastic Gradient Descent}
\acro{ROC}[ROC]{Receiver Operating Characteristic}
\acro{MFV}[MFV]{Multifaceted Feature Visualization}
\acro{t-SNE}[t-SNE]{t-Distributed Stochastic Neighbor Embedding}
\acro{ReLU}[ReLU]{Rectified Linear Unit}
\acro{LRN}[LRN]{Local Response Normalization}
\acro{PAI}[PAI]{PhotoAbsorption and Ionization}
\acro{LHC}[LHC]{Large Hadron Collider}
\acro{MWD}[MWD]{Micro-Well Detector}
\acro{LAT}[LAT]{Large Area Telescope}
\acro{GPU}[GPU]{Graphics Processing Unit}
\acro{GCAM}[Grad-CAM]{Gradient-weighted Class Activation Mapping}
\acro{ILSVRC}[ILSVRC]{ImageNet Large Scale Visual Recognition Competition}
\acro{NASA}[NASA]{National Aeronautics and Space Administration}
\acro{SoC}[SoC]{System on a Chip}
\end{acronym}


\begin{frontmatter}

\title{Event Selection and Background Rejection in Time Projection Chambers Using Convolutional Neural Networks and a Specific Application to the AdEPT Gamma-ray Polarimeter Mission}

\author[mac_medphys]{Richard L. Garnett}
\ead{garnetri@mcmaster.ca}

\author[mac_medphys,mac_phys]{Soo Hyun Byun}
\author[mac_medphys,mac_phys]{Andrei R. Hanu}
\ead{hanua@mcmaster.ca}
\author[nasa]{Stanley D. Hunter}

\address[mac_medphys]{Radiation Sciences Graduate Program, McMaster University, Tandem Accelerator Building, 1280 Main St. W, Hamilton, ON, L8S 4K1, Canada}
\address[mac_phys]{Department of Physics and Astronomy, McMaster University, Hamilton, ON, L8S 4K1, Canada}
\address[nasa]{NASA Goddard Space Flight Center, Greenbelt, MD 20771, USA}



\begin{abstract}

The Advanced Energetic Pair Telescope gamma-ray polarimeter uses a time projection chamber for measuring pair production events and is expected to generate a raw instrument data rate four orders of magnitude greater than is transmittable with typical satellite data communications. \texttt{GammaNet}, a convolutional neural network, proposes to solve this problem by performing event classification on-board for pair production and background events, reducing the data rate to a level that can be accommodated by typical satellite communication systems. In order to train \texttt{GammaNet}, a set of 1.1x10$^6$ pair production events and 10$^6$ background events were simulated for the Advanced Energetic Pair Telescope using the \texttt{Geant4} Monte Carlo code. An additional set of 10$^3$ pair production and 10$^5$ background events were simulated to test \texttt{GammaNet}'s capability for background discrimination. With optimization, \texttt{GammaNet} has achieved the proposed background rejection requirements for Galactic Cosmic Ray proton events. Given the best case assumption for downlink speeds, signal sensitivity for pair production ranged between 1.1$\pm$0.5\% to 69$\pm$2\% for 5 and 250 MeV incident gamma rays. This range became 0.1$\pm$0.1\% to 17$\pm$2\% for the worst case scenario of downlink speeds. The application of a feature visualization algorithm to \texttt{GammaNet} demonstrated decreased response to electronic noise and events exiting or entering the frame and increased response to parallel tracks that are close in proximity. \texttt{GammaNet} has been successfully implemented and shows promising results.

\end{abstract}


\end{frontmatter}



\acresetall

\section{Introduction}
\label{Introduction}

Recent advances in machine learning and computer vision have led to astonishing improvements in image classification performance \cite{hoo2016deep,szegedy2016inception,iandola2016squeezenet}, where algorithms estimate the likelihood that an input image belongs to a set of labels that describe features contained within the image. Current state of the art algorithms perform with around 1.3\% top-5 error \footnote{Where the top-5 error is determined by the fraction of test images for which the correct label is not among the five labels considered most probable by the algorithm.} \cite{touvron2020fixing}. These results were demonstrated on test sets of images from the \ac{ILSVRC} \cite{ILSVRC15} which contain images belonging to 10$^3$ different classes.

The application of machine learning to event classification in radiation detection is a natural progression of the field given that radiation detectors produce highly structured signals. These signals are often dependent on the nature of interacting radiation, and the type of interaction undergone. High energy physics projects such as the Large Hadron Collider have utilized machine learning applications for event classification \cite{baldi2014searching, baldi2016jet}. There has also been implementations of machine vision for image classification in radiation imaging detectors using \acp{CNN} to classify neutrino interactions at Fermilab and the Ash River Laboratory \cite{NeutrinoCNN}. 

The \ac{CNN} application explored in this work has been developed for the event classification of images generated from a large (8 m$^3$) \ac{TPC} being designed for the \ac{AdEPT} \cite{hunter2014pair}, a mission to measure medium-energy gamma-ray polarimetry. The design details of \ac{AdEPT} are discussed in detail in \cite{hunter2014pair}, and briefly summarized in \cref{subsection:Instrument}.

\subsection{The \ac{AdEPT} Instrument}
\label{subsection:Instrument}

Astrophysical gamma rays are a means to probe the most extreme non-thermal processes in the Universe and their study provides valuable insight into the fundamental physics and structure of the most powerful natural particle accelerators. Most studies of astrophysical gamma rays have been in the $\sim$20 MeV to 300 GeV energy range, using measurements from the AGILE \cite{tavani2009agile} and Fermi \cite{atwood2009large} space telescopes. However, neither instrument was optimized for polarization sensitivity or observations in the medium energy ($\sim$0.1--200 MeV) band, where many astrophysical objects exhibit unique behavior. The medium energy gamma-ray band has so far proven difficult to study due to competing photon interactions, namely Compton scatter and pair production. Each of these interactions generate different signatures, and the manner in which polarization information is gathered consequently requires differing algorithms and instrumentation \citep{hunter2014pair, lei1997compton, forot2008polarization}. The optimization of a detector for both pair production and Compton scatter interactions on-board a satellite is prohibitive. The challenge is further exacerbated by the \ac{GCR} background, which is an extragalactic source of charged atomic nuclei at extremely high kinetic energy. The \ac{GCR} background cannot be effectively shielded for on satellites given their high kinetic energy, which can extend to several TeV per nucleon. In addition, the fluence of \ac{GCR} particles exceeds the astrophysical gamma-ray flux by approximately four orders of magnitude.

Next-generation telescopes are being developed with the goal of characterizing the complete signature of gamma rays including their direction, energy, arrival time, and polarization. The most promising space missions (\acs{AdEPT} \cite{hunter2014pair}, HARPO \cite{bernard2014}, and SMILE-I/II \cite{takada2011observation,ueno2012development}) proposed to explore the gamma-ray sky in the medium energy range are based on low-density  gaseous \ac{TPC} technologies that enable precise, three-dimensional tracking of particle interactions. 

The \ac{AdEPT} mission is one such medium energy gamma-ray polarimeter. The science data for \ac{AdEPT} will consist of pair production interactions, with a background composed primarily of \ac{GCR} and Compton scatter interactions. Compton scatter, though a photon interaction of interest for characterizing the medium energy gamma-ray spectra, is considered background for the \ac{AdEPT} mission. Compton scatter is considered background because the \ac{AdEPT} instrument is not designed to measure polarization for this interaction. The \ac{AdEPT} \acp{TPC} takes advantage of the \ac{MWD} technology augmented with the negative ion drift technique \cite{martoff2005negative} to construct an instrument with the largest volume that can be accommodated in the rocket fairings currently available to MIDEX missions, 8 m$^3$. The active gas volume of the \ac{TPC} is bounded on the top and bottom faces by an array of \acp{MWD} defining the 400 $\upmu$m X- and Y-coordinate spatial resolution of the \ac{TPC} \cite{hunter2014pair}. The uniform electric field in the active volume provides a constant ionization charge drift velocity. Measurement of a relative arrival time of the signals on the detector strips provides the third, Z-coordinate. The use of the negative ion drift technique in the \ac{AdEPT} \ac{TPC} design \cite{hunter2014pair} effectively reduces electron drift diffusion in the gas, making possible drift distances up to 1 m. With the applied electric field, ionization charge can traverse the Z dimension of the detector within a maximum of 50 ms. 

The use of the negative ion drift technique precludes the use of an anti-coincidence system, as used in HARPO \cite{bernard2014}, resulting in large raw data rates. This requires an alternative on-board processing approach for discrimination of \ac{GCR} tracks and gamma-ray interactions. The 8 m$^3$ version of \ac{AdEPT} is estimated to produce an uncompressed data rate of $\sim$16 Gbps, far too large for current satellite communication. Currently the Fermi Large Area Telescope mission \cite{atwood2009large,meegan2009fermi,cameron2012fermi} achieves an average science data downlink of 1.5 Mbps, while planned communications methods aim to achieve an average 50 Mbps downlink \cite{robinson2018terabyte}. The range of possible average downlinks leaves two to four orders of magnitude difference between the raw data rate and communications data rate for the \ac{AdEPT} mission. Our proposed solution is to use computer vision feature recognition algorithms running on-board the spacecraft in real time to discriminate gamma-ray interactions of interest from the abundant \ac{GCR} background. The desired outcome for the algorithm is to perform event classification within 50 ms with a background rejection rate between 99.99\% and 99.69\%, which would reduce the raw data rate to one which can be accommodated by satellite downlink. The hardware to be used for \ac{AdEPT} has not yet been chosen, though commercial solutions are available that offer enough computing power for \texttt{GammaNet}. One such solution is Innoflight's Compact Flight Computer 500, which is radiation tolerant up to 30 krad, and is space rated. Additionally, \ac{NASA} is investigating the suitability of \ac{SoC} solutions available from NVIDIA \cite{powell2018commercial}.

In this paper we explore \texttt{GammaNet}, a \ac{CNN} trained on simulated images from a high resolution gaseous \ac{TPC}, and its performance in classifying gamma-ray events on images contaminated with  a \ac{GCR} background. To evaluate the performance of \texttt{GammaNet}, we performed a \ac{ROC} analysis \cite{fawcett2006introduction} to assess how the background rejection threshold influences the specificity and sensitivity of the classifier. Specificity is defined as the rate at which negative events are correctly classified as negative. Sensitivity is defined as the rate at which positive events are correctly classified as positive. The result of our \ac{ROC} study demonstrated that \texttt{GammaNet} can reliably achieve the proposed background rejection rates of between 99.99\% and 99.69\%. At these rates of background rejection, \texttt{GammaNet} correctly classifies between 10$\pm$1\% and 52$\pm$2\% of pair production images over the energy range of interest. The simulation used for generating training and testing data, as well as the architecture and training protocol for GammaNet, are described thoroughly in \cref{section:MCSim,section:GammaNet}. An analysis of \texttt{GammaNet's} performance and failures is presented in \cref{section:Results}. Observations are presented in \cref{section:GradCAM} for the features utilized by \texttt{GammaNet} classifying the simulation images of \ac{AdEPT}.


\begin{figure*}[b] 		
	\centering
	\captionsetup{justification=centering}	
	\begin{subfigure}[t]{0.32\textwidth}
		\centering
		\includegraphics[width=\textwidth]{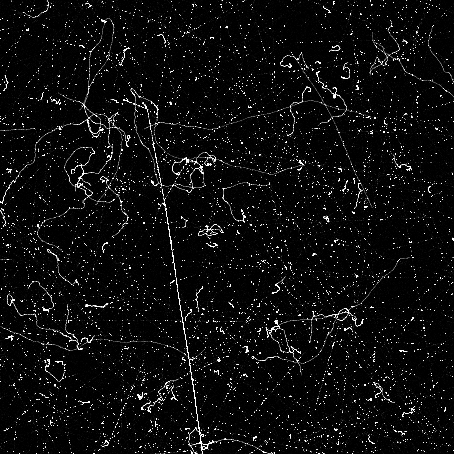}
		\caption{}
		\label{subfig:GCRAlone}
	\end{subfigure}
	~
	\begin{subfigure}[t]{0.32\textwidth}
		\centering
		\includegraphics[width=\textwidth]{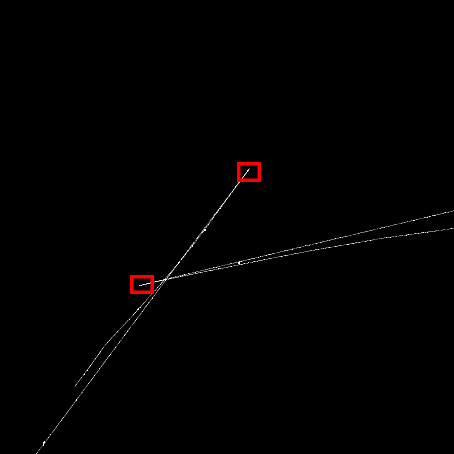}
		\caption{}
		\label{subfig:PPAlone}
	\end{subfigure}
	~	
	\begin{subfigure}[t]{0.32\textwidth}
		\centering		
		\includegraphics[width=\textwidth]{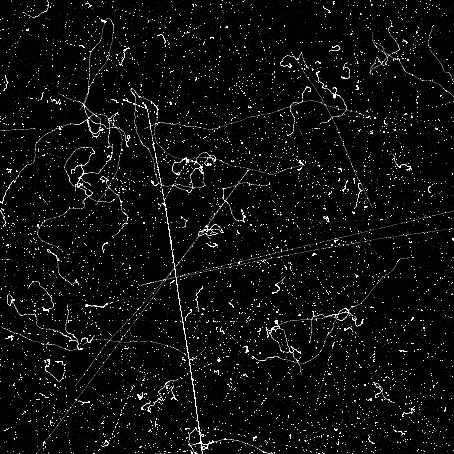}
		\caption{}
		\label{subfig:PPGCRCombined}
	\end{subfigure}
	\caption{XZ projection of the sensitive volume of the \ac{AdEPT} simulation. a) \ac{GCR} background image containing several proton tracks with added electronic noise. b) gamma-ray image, containing two pair production events with the vertices outlined in red for illustrative purposes. c) Combination image that would be used for training and testing \texttt{GammaNet}. These simulation images have had their contrast adjusted for better viewing in this paper.}
	\label{fig:SimulationImage}
\end{figure*}

\section{Monte Carlo Simulations of \ac{AdEPT}}
\label{section:MCSim}

The \ac{AdEPT} raw \ac{TPC} data consists of two orthogonal projections, XZ and YZ, of the tracks in the active gas volume. The simulation of the response and readout of the \ac{AdEPT} instrument was carried out using the \texttt{G\textsc{eant}4} Monte Carlo toolkit \citep{agostinelli2003geant4,allison2006geant4}. The application, which is named \texttt{G4AdEPTSim} \cite{HanuSim}, simulates the passage of \ac{GCR} protons and gamma rays through an active volume filled with 1.5 atmospheres of Ar and CS$_2$ at a temperature of 293 degrees K with a sub-scale size of 25x25x25 cm$^3$, and full-scale size of 8 m$^3$. 

The use of the sub-scale volume was to determine what level of downsampling was viable for use in \texttt{GammaNet}, and subsequently the full-scale volume was used to determine the performance of \texttt{GammaNet}. In this work, downsampling is the process of taking an N x N region of the image, averaging it, and applying it to a single pixel in the output. This scales down the image by a factor of N$^2$, which is necessary for this work because the time to train and run classification for any \ac{CNN} is strongly correlated to the image size passed to it. The full size \ac{AdEPT} \ac{TPC} will produce images of 5000 x 5000 pixels, which would be prohibitively slow in terms of both training and time to classification during operation.

The physics included in the simulation account for the different types of interactions between source particles and the Ar gas. These include hadronic physics for the interaction of \ac{GCR} protons, electromagnetic physics for the interaction of gamma rays and electrons, and photo-absorption ionization model to accurately model the primary ionization and energy loss of relativistic charged particles in low density media. \texttt{G4AdEPTSim} produces the ideal response of AdEPT, reporting the number of ionization electrons, their X-, Y-, and Z-coordinates, and the energy deposited in the active volume by a single incident particle. 

\ac{AdEPT} is proposed for launch into a low-Earth orbit with a 550 km altitude and a 28 degree inclination. The background environment in such an orbit is well-known and consists predominantly of \acp{GCR}, cosmic diffuse radiation, atmospheric \linebreak gamma-ray emissions, reactions induced by albedo neutrons, and background produced by satellite materials activated by fast protons and alpha particles \citep{badhwar1997radiation, benton2001space, zhou2006radiation, weidenspointner2000cosmic, henry1999diffuse}. In the ~0.1 to 200 MeV energy range, the instrument background is dominated by charged particles in the Van Allen belt impinging on the spacecraft, cosmic diffuse radiation, and atmospheric gamma-ray emissions. 

\texttt{G4AdEPTSim} models the simulated events using a spherical volume source of radius 22 cm for the sub-scale version, and 1.73 m for the full-scale version, which is concentric with the active volume. The arrival direction of the simulated particles is isotropically distributed on a sphere, producing a uniform distribution of the source particles within the sphere. For this work, the background component consisted of only \ac{GCR} protons with the energy spectrum from the Space Environment Information System for the expected \ac{AdEPT} orbital conditions. \ac{GCR} protons were selected as the background because they comprise the majority of the \ac{GCR} fluence. Astrophysical sources of gamma rays simulated with single energies ranging from 5--250 MeV were generated using the same source geometry as background.

Each simulation run of the full-scale \ac{AdEPT} instrument contained 375 incident \ac{GCR} protons for signal and background, with an additional two incident gamma rays for signal. The sub-scale simulation runs consisted of five incident \ac{GCR} protons or two incident gamma rays to account for the reduced surface area relative to the full-scale instrument. The number of incident particles were chosen in each case to fit the expected number of primary tracks, given the \ac{AdEPT} instrument parameters \citep{hunter2014pair}, within the 50 ms collection window. There are two incident gamma rays for both simulations because the anticipated pair production rate in the full size simulation is less than one, although there is still the probability of two pair production events occurring within one collection window. For the gamma ray events an interaction is forced if the path intersects the sensitive volume of the detector. This interaction type is forced as pair production for the signal events, and Compton scatter for the investigation of \texttt{GammaNet}’s sensitivity to Compton scatter. The source geometry allowed for the possibility of particles to miss the active volume, but results were only recorded if at least one particle interacted with the active volume. The source geometry used allows for a varying number of tracks to be recorded from each simulation run, although the number of simulated particles was constant between runs. 

Per simulation run the number of ionization electrons in 400 x 400 x 400 $\upmu$m$^3$ voxels was recorded, corresponding to the nominal resolution of the \ac{AdEPT} instrument. The number of ionization electrons in each voxel is then projected onto the XZ and YZ planes to generate images. To emulate the response of the \ac{AdEPT} detector, electronic noise was added to the signal output for each set of images. The addition of electronic noise was performed by adding a randomly generated number of electrons, from a normal distribution with standard deviation of two and a mean of zero, to each pixel of an image. In addition to electronic noise, background events were added to every gamma-ray image in the form of \ac{GCR} protons. To do the background event addition, \ac{GCR} proton images were generated with electronic noise and gamma-ray images without. Each gamma-ray image then had a unique \ac{GCR} image added to it. Gamma-ray images were generated without the addition of electronic noise to ensure \ac{GCR} images and the composite gamma-ray images would have a constant amount of electronic noise. \cref{fig:SimulationImage} shows an example of the process used for generating the pair production data set, where an image containing two pair production events is added to a background \ac{GCR} image with two tracks. 
 

Correctly labeled image sets were generated from these simulations for both training and testing of \texttt{GammaNet}. The training image sets contained 1.1x10$^6$ pair production images and 10$^6$ background \ac{GCR} proton images. The testing image sets contained 1.5x10$^3$ pair production images, 1.5x10$^3$  Compton scatter images, and 10$^6$  background \ac{GCR} images. The Compton scatter images were included in testing, but not training, as an additional source of background. \texttt{GammaNet} was found to be less sensitive to the Compton scatter images than pair production. The intuition when applying a \ac{CNN} to this classification problem was that the \ac{CNN} would be able to pick up on the discerning characteristic of pair production events compared to \ac{GCR} proton tracks. These pair production signatures are further discussed in the results, \cref{section:Results,section:GradCAM}.

\section{GammaNet}
\label{section:GammaNet}

\texttt{GammaNet} was inspired by the successes of a \ac{CNN} designed for classification of neutrino interaction events in the NOvA experiment at Fermilab \cite{ayres2005nova}. The \ac{CNN} showed an increased performance compared to the state-of-the-art algorithms currently deployed for classification of neutrino events at Fermilab. Specifically, there was a relative increase of 40\% sensitivity for electron neutrino signals, going from 35\% to 49\% \cite{NeutrinoCNN}. However, the \ac{AdEPT} instrument does not require as much information about the background radiation as the NOvA detector, allowing \texttt{GammaNet} to classify to two classes as opposed to NOvA's 13.
\texttt{GammaNet} produces a probability that the input image from \ac{AdEPT} contains a pair production event, which, above a certain threshold, will result in a positive signal, and below will produce a negative signal. This simplicity allows for faster classification with an unsophisticated architecture.

\subsection{GammaNet Architecture}

\begin{figure}[b]
	\centering
	\captionsetup{justification=centering}	
	\begin{subfigure}[t]{0.48\textwidth}
		\centering
		\includegraphics[width=\textwidth]{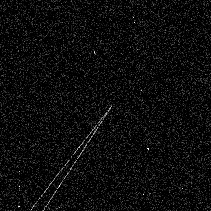}
		\caption{}
		\label{subfig:XZProjection}
	\end{subfigure}
	~
	\begin{subfigure}[t]{0.48\textwidth}
		\centering		
		\includegraphics[width=\textwidth]{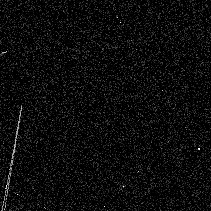}
		\caption{}
		\label{subfig:YZProjection}
	\end{subfigure}
	\caption{XZ and YZ projections of the same event generated in the sub-scale simulation, with a downsampling rate of 3. In a), the XZ projection, a well separated pair production track is shown in the lower half of the image. In b), the YZ projection, an overlapping pair production track is shown in the lower left of the image.}
	\label{fig:Overlap}
\end{figure}

The XZ and YZ projection images that the \ac{AdEPT} \ac{TPC} produce are used as the input of two identical instances of \texttt{GammaNet} for classification. The classifications of the two projections are then compared using a boolean operation, where if either projection produced a positive signal, the event was determined to be positive. Comparing both projections helps reduce errors associated with the positron and electron tracks overlapping in a projection, which would appear as a singular track. Having the two orthogonal projections ensures that this overlap is avoided in at least one of the images provided to \texttt{GammaNet}, avoiding misclassification of pair production events. An example of this issue is shown in \cref{fig:Overlap}, where in the first projection, the two tracks from the pair production are well separated, and the alternative projection shows them overlapping to an extent. The architecture of \texttt{GammaNet} is presented in \cref{subsection:appendix-architecture}.

\texttt{GammaNet}'s architecture is an adaptation of \texttt{GoogLeNet} \cite{szegedy2015going} with modifications needed for reduced time to classification and the stringent background rejection requirements of \ac{AdEPT}. To reduce time to classification, the overall network size was truncated by utilizing only one inception module, where an inception module is a network in network design created by Google \cite{szegedy2015going}. \cref{table:Accuracies} lists the results from \texttt{GammaNet} when operating with a threshold of 0.5 for classification of background and pair production when differing the number of inception modules. From these results it is shown that the highest \ac{GCR} background rejection rate was achieved with a single inception module. 

This result is counterintuitive given that it is generally accepted that increasing the depth of a \ac{CNN} increases its classification performance. The idea of increased classification performance with increased depth in \acp{CNN} is based on classification of real life images, such as those presented in the \ac{ILSVRC}, not radiation interactions as is with \texttt{GammaNet}. Given that radiation interaction images contain significantly less information than a real life image, it is our interpretation that a relatively shallow \ac{CNN} would be able to learn the features required to classify those images successfully. Our expectation is that lack of complexity in our images allows for \texttt{GammaNet} to perform optimally with a comparatively shallow \ac{CNN}, with only marginal differences in performance for increased network depth, as is shown in \cref{table:Accuracies}. 

\newcolumntype{C}[1]{>{\centering\arraybackslash}p{#1}}
\begin{table}[h]
\caption{Tabulated results of \texttt{GammaNet} pair production sensitivity and background rejection rate for differing numbers of inception modules. Pair Production sensitivity reported as highest of the 5--250 MeV energy sets whereas background rejection rate was calculated from only one set.}  
\begin{tabular*}{0.9\textwidth}{*3{C{0.25\textwidth}}}
    \toprule
     Number of Inception Modules  & Pair Production Sensitivity (\%) & Background \ac{GCR} Rejection Rate (\%) \\
    \midrule
    1  &  \textbf{93.17} & \textbf{96.30}\\
    2  &  94.28 & 93.91 \\
    3  &  93.47 & 96.10 \\
    \bottomrule
    \label{table:Accuracies}
\end{tabular*}
\end{table}

\subsection{Training}

\begin{figure}[bh!]
	\centering
	\captionsetup{justification=centering}
	\includegraphics[width=0.9\textwidth]{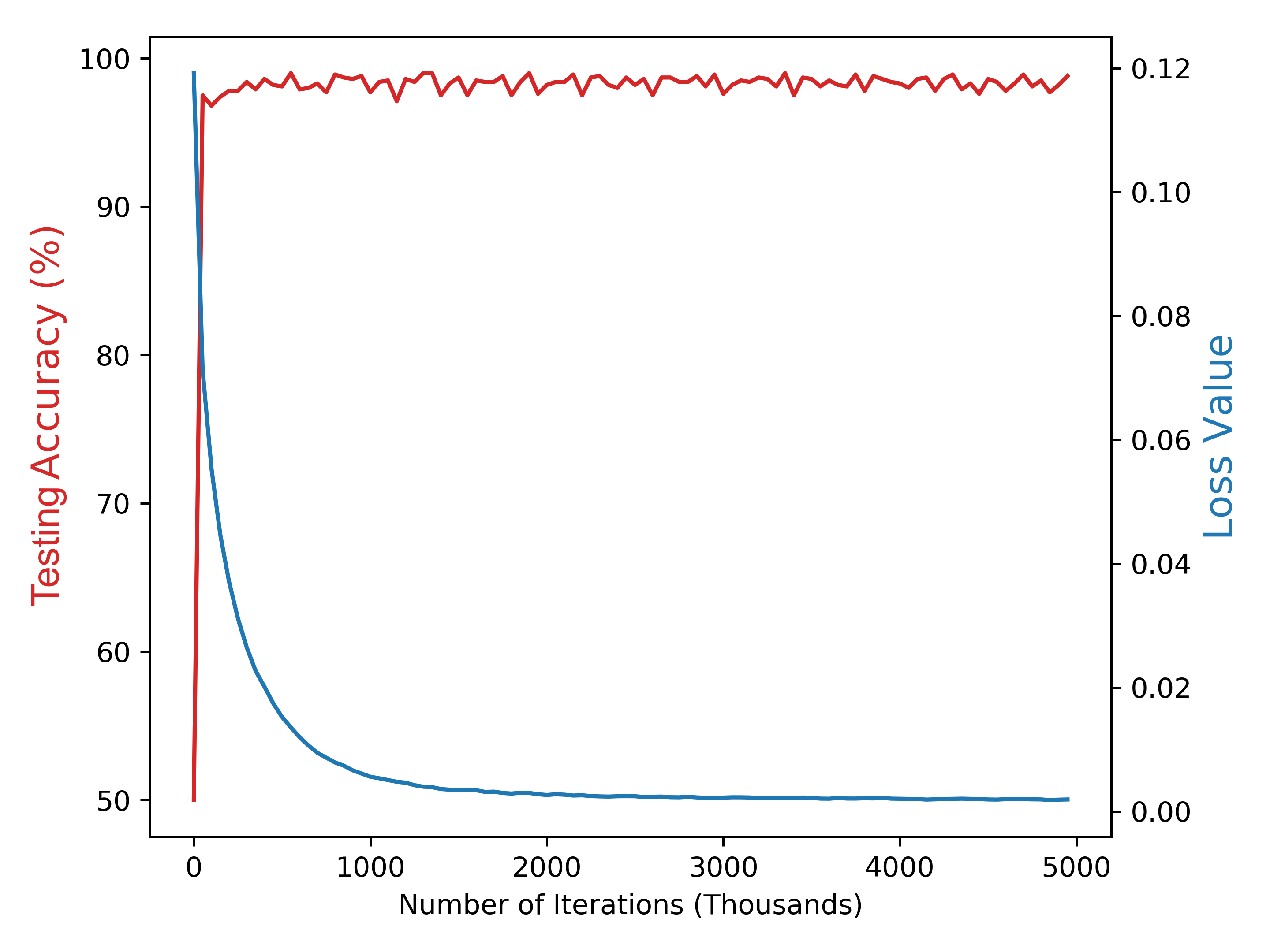}
	\caption{Graph of the training results for \texttt{GammaNet} with 1 inception module. This training data was generated with the sub-scale simulation, with a downsampling rate of 3x. The left axis contains the accuracy of \texttt{GammaNet} on the validation data set, and the right axis contains the loss value averaged over every 50k training iterations.}
	\label{fig:Training}
\end{figure}

The training process for \texttt{GammaNet} involves passing a simulation image through it, after which the parameters of each layer in the network are updated based on the negative gradient of that output with respect to each parameter. Training is continued until the network converges on a steady state of accuracy with respect to a testing data set that is separate from the training data. The training procedure is governed by a handful of parameters, called hyperparameters, used by \texttt{NVCaffe} to determine how training is carried out \cite{jia2014caffe}. The hyperparameters used for training \texttt{GammaNet} can be seen in \cref{subsection:Appendix-hyperparams}. The results of training are shown in \cref{fig:Training}, and demonstrates that \texttt{GammaNet} converges upon a solution quickly while training. Training was continued for 5 million iterations for each version of \texttt{GammaNet}, and to 2 million iterations for the \texttt{VGG16} \cite{simonyan2014very} architecture. The training graphs of subsequent networks were omitted for the sake of brevity, though each network reached similar results to \cref{fig:Training}.

\section{Results}
\label{section:Results}

The final layer of \texttt{GammaNet}, \cref{fig:NetworkDiagram} m, outputs the probability that a given input image contains a gamma-ray pair production signal or is purely background. To analyze the performance of \texttt{GammaNet} as a binary classifier a \ac{ROC} analysis \cite{fawcett2006introduction}, which determines a classifier's specificity and sensitivity at different threshold values, was conducted. In the \ac{ROC} algorithm, the list of classification outputs produced by \texttt{GammaNet} for the image set is sorted by decreasing value of probability for the pair production event class. The threshold value is then iterated through the list of pair production class probabilities. For each iteration the classification probability for pair production produced by \texttt{GammaNet} in response to the input image is compared to the threshold. If the classification probability for pair production is lower than the threshold, the image is classified as background. If the classification probability for pair production is above threshold, the image is classified as pair production. Utilizing the threshold for classification allows for an event classified as pair production to be either a true positive or false positive event. The number of true positive and false positive events is then tallied and normalized to the number of images in the set to generate the true and false positive rates for each threshold value. 

\begin{figure}[hb!]
	\centering
	\captionsetup{justification=centering}
	\begin{subfigure}[t]{1\columnwidth}
	\centering
	\includegraphics[width=0.95\columnwidth]{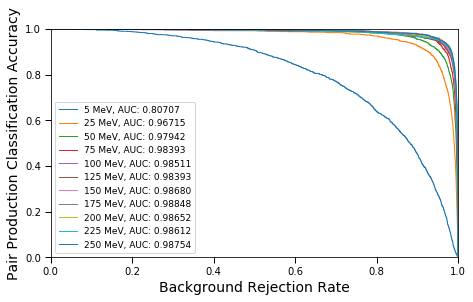}
	\caption{}
	\label{subfig:PP200MeVROCsub1}
	\end{subfigure}
	\\
	\begin{subfigure}[t]{1\columnwidth}
	\centering
	\includegraphics[width=0.95\columnwidth]{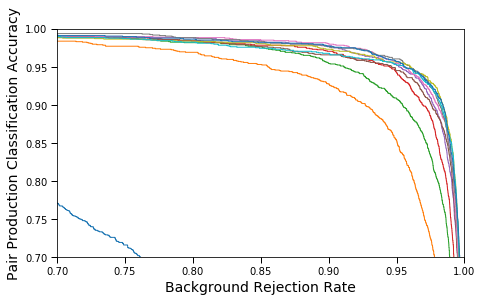}
	\caption{}
	\label{subfig:PP200MeVROCsub2}
	\end{subfigure}
    \caption{a) The \ac{ROC} curves generated using the described algorithm for each pair production data set as classified by \texttt{GammaNet}, using 11x downsampled images. The AUC is provided in the legend for each incident gamma-ray energy, with an area of 1 being a perfect classifier, and 0.5 being random selection. b) A subsection of \cref{subfig:PP200MeVROCsub1} is presented to display the nuanced features of the plot.}
	\label{fig:PP200MeVROC}
\end{figure}

The \ac{ROC} plot provides a graphic representation of the classifier's response to threshold levels for true positive and false positive rates. The \ac{ROC} curve generated for \texttt{GammaNet} is shown in \cref{fig:PP200MeVROC}. Area Under the \ac{ROC} Curve (AUC) in \cref{fig:PP200MeVROC} ranges from 0.807 to 0.988 depending on incident gamma-ray energy, demonstrating the general level of performance of \texttt{GammaNet} as a binary classifier. The individual points on \cref{fig:PP200MeVROC} show the sensitivity to pair production of \texttt{GammaNet} at a given background rejection rate, which can be used to determine what threshold to run \texttt{GammaNet} at to satisfy the requirements of \ac{AdEPT} for background rejection rates.

Using the \ac{ROC} analysis, it is possible to investigate the performance of \texttt{GammaNet} with respect to the rate of downsampling used on the simulation data. Downsampling results in the projection images being reduced by a factor of N$^2$, which significantly reduces the time taken to train and perform classification with \texttt{GammaNet}. The impacts of downsampling on signal sensitivity are displayed in \cref{table:DownSampling}, where downsampling rates between 1--11 were investigated using the sub-scale simulation. When downsampling by 1 the voxel size is maintained at 400 x 400 x 400 $\upmu$m$^3$, and when downsampling by 11 the voxel size is reduced to 4.4 x 4.4 x 4.4 mm$^3$. 

From the results in \cref{table:DownSampling}, it is shown that any amount of downsampling outperforms the alternative of no downsampling, with a decrease in signal sensitivity for increasing downsampling rates. This is of benefit to \texttt{GammaNet} for both the increase in sensitivity and that at higher downsampling rates \texttt{GammaNet} can perform image classification in less time. This increase in sensitivity for any amount of downsampling is expected to be due to the original images containing discontinuities in the ionization tracks from the pair production events, this is reduced or entirely removed when downsampling the image. This gain in sensitivity is then diminished at greater degrees of downsampling as the higher rates of downsampling reduce the ability to distinguish both arms of the pair production tracks. Given the large image size generated by the full-scale simulation, a downsampling rate of 11 was used for the remainder of the work when utilizing the full-scale simulation. Training of \texttt{GammaNet} on the full-scale simulation data at a downsampling rate of 11 took 30 days of compute time, proving investigating \texttt{GammaNet's} performance on lower downsampling rates with the full-scale simulation data to be prohibitively time consuming.

\begin{table}[h!]
\caption{Pair Production sensitivity for \texttt{GammaNet} and VGG16 at varying background rejection rates corresponding to anticipated downlink speeds. Performance comparison results were generated using the sub-scale simulation data, with a downsampling rate of 3.}  
\begin{tabular*}{0.9\columnwidth}{*3{C{0.25\columnwidth}}}
    \toprule
     Background Rejection Rate (\%) & \texttt{GammaNet} Pair Production Sensitivity (\%) & VGG16 Pair Production Sensitivity (\%) \\
    \midrule
    99.990$\pm$0.002 &  65$\pm$2  & 28$\pm$2 \\
    99.97$\pm$0.003   &  73$\pm$2 & 38$\pm$2 \\
    99.94$\pm$0.005   &  78$\pm$2 & 46$\pm$2 \\
    99.87$\pm$0.007   &  84$\pm$2 & 57$\pm$2 \\
    99.81$\pm$0.009   &  87$\pm$2 & 61$\pm$2 \\
    99.75$\pm$0.01   &  89$\pm$2 & 64$\pm$2 \\
    99.69$\pm$0.01   &  90$\pm$1 & 66$\pm$2 \\
    \bottomrule
    \label{table:VGG16}
\end{tabular*}
\end{table}

\begin{table*}[t]
    \begin{tabular}{ccccccccc}
    & \multicolumn{2}{c}{} &  \multicolumn{6}{c}{Downsampling Rate} \\ \cline{4-9}
    & \multicolumn{3}{r}{1} & \multicolumn{1}{c}{3} & \multicolumn{1}{c}{5} & \multicolumn{1}{c}{7}
    & \multicolumn{1}{c}{9} & \multicolumn{1}{c}{11} \\ \midrule
    & \makecell{Data Rate Limit\\ (Mbps avg.)} & \makecell{Background Rejection\\ Rate (\%)} 
    & \multicolumn{5}{c}{\hspace{10ex}Signal Sensitivity (\%)}& \\ \midrule
    & 1.5 & 99.99$\pm$0.002 & 43$\pm$2 & 65$\pm$2 & 50$\pm$2 & 41$\pm$2 & 40$\pm$2 & 28$\pm$2 \\
    & 5   & 99.97$\pm$0.003 & 47$\pm$3 & 73$\pm$2 & 65$\pm$2 & 54$\pm$2 & 49$\pm$2 & 45$\pm$2 \\
    & 10  & 99.94$\pm$0.005 & 54$\pm$3 & 78$\pm$2 & 73$\pm$2 & 62$\pm$2 & 56$\pm$2 & 56$\pm$2 \\
	& 20  & 99.87$\pm$0.007 & 64$\pm$2 & 84$\pm$2 & 81$\pm$2 & 74$\pm$2 & 67$\pm$2 & 66$\pm$2 \\
	& 30  & 99.81$\pm$0.009 & 68$\pm$2 & 87$\pm$2 & 84$\pm$2 & 78$\pm$2 & 72$\pm$2 & 71$\pm$2 \\
	& 40  & 99.75$\pm$0.01 & 71$\pm$2 & 89$\pm$2 & 86$\pm$2 & 81$\pm$2 & 75$\pm$2 & 74$\pm$2 \\
	& 50  & 99.69$\pm$0.01 & 74$\pm$2 & 90$\pm$1 & 87$\pm$2 & 83$\pm$2 & 78$\pm$2 & 77$\pm$2 \\ \bottomrule
    \end{tabular}
    \caption{Pair production sensitivity of \texttt{GammaNet}, for sub-scale simulation images, given the desired background rejection rate with differing factors of downsampling. The data rate limits are sampled between the proposed minimum and maximum as described in \cref{Introduction}. The background rejection rates listed are calculated by using the ratio of the raw data rate and the data rate limit, assuming the signal is approximately entirely background. Each data set was generated from the sub-scale simulation, using the given downsampling rate. \texttt{GammaNet} was then trained and tested on those data sets. The reported pair production signal sensitivities are the average sensitivity for the energies simulated. Error was calculated using binomial statistics with a 95\% confidence interval.}
	\label{table:DownSampling}
\end{table*}

A cursory investigation between the performance of \texttt{GammaNet} relative to other neural network architectures was performed. In this investigation another neural network architecture was chosen, VGG16 \cite{simonyan2014very}, given it outperformed GoogLeNet in the \ac{ILSVRC}. VGG16 was trained in the exact same manner as \texttt{GammaNet} and the performance of the two networks were compared. The performance comparison between \texttt{GammaNet} and VGG16 was carried out with a downsampling rate of 3, and with data produced from the sub-scale simulation. \cref{table:VGG16} provides the results from each network when classifying the sub-scale simulation data, with \texttt{GammaNet} shown to largely outperform VGG16 over the entire range of background rejection rates investigated. This result is not what would be anticipated given that VGG16 outperformed GoogLeNet in the \ac{ILSVRC} competition, but the task of event classification for \ac{AdEPT} utilizes more sparse images. These results imply that \texttt{GammaNet} is more well suited for classifying background images than is VGG16, which ultimately is the primary task of \texttt{GammaNet} for \ac{AdEPT}.

\begin{figure}[bh!]
	\centering
	\includegraphics[width=0.95\columnwidth]{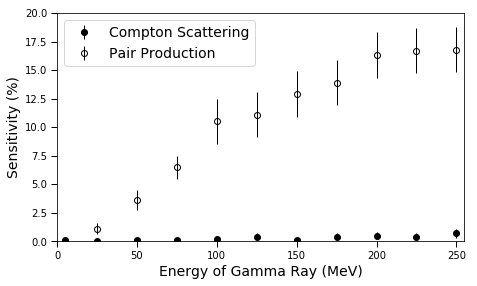}
	\caption{Plot of the sensitivity for Compton scatter and pair production image sets as the energy of the incident gamma ray varies, using a downsampling rate of 11 on the full-scale simulation. These sensitivities were calculated using a threshold value that generated a 99.990$\pm$0.002\% background rejection rate. Errors were calculated using binomial statistics with a 95\% confidence interval.}
	\label{fig:ROCwE}
\end{figure}

The performance of \texttt{GammaNet} when classifying Compton scatter events was of interest as well given that it is the main gamma-ray interaction contributing to background in the \ac{AdEPT} instrument, and the similarity in track structure compared to pair production. The rate of misclassification for Compton scatter events as pair production events provides information about the features that \texttt{GammaNet} uses for classifying the input. The main differentiation between the pair production and Compton scatter tracks is the presence of only a singular track for Compton scatter and the absence of the vertex from pair production. Support for the importance of these features for classification is shown in \cref{fig:ROCwE}. As the incident gamma-ray energy increases, so too does the signal sensitivity for pair production. The increase in signal sensitivity is due to the increased energy of the positron-electron pair producing more linear tracks, closer in proximity, and with more distinct vertices. This is supported by the negligible increase in signal sensitivity for Compton scatter events.

Due to the large raw data rate and the limits of satellite communications, it is required to achieve a background rejection rate of 99.99\% to 99.69\% in order for the data to be transmitted. To achieve this background rejection rate, the threshold for a pair production event classification has to be set quite high, which results in a number of pair production events being misclassified as background events. \cref{table:accuracies} shows the average rate at which \texttt{GammaNet} classifies pair production and Compton scatter events as a positive event, given different background rejection rates. These results were generated using the full-scale simulation with a downsampling rate of 11. The classification accuracies were averaged over the energies simulated for pair production and Compton scatter. It is shown in \cref{table:accuracies} and \cref{fig:ROCwE} that at the proposed 99.99\% background rejection rate, we obtain a pair production sensitivity between 0.1$\pm$0.1\% and 17$\pm$2\%, depending on incident photon energy, with an average of 10$\pm$1\%. For the best case scenario of 99.69\% background rejection, the signal sensitivity increases to a range of 1.1$\pm$0.5\% to 69$\pm$2\%, again depending on incident photon energy, with an average of 52$\pm$2\%. In both cases, the sensitivity to Compton scatter is quite small, which is beneficial for the mission due to Compton scatter representing background for the \ac{AdEPT} mission. The relatively low sensitivity to pair production events at low energy will reduce the effectiveness of the instrument, but this impact can be mitigated during mission design by implementing image compression, where these calculations were done assuming no compression.

\begin{table}[h!]
\caption{Pair production and Compton scatter sensitivity at varying background rejection rates corresponding to anticipated downlink speeds. The \ac{GCR} proton background rejection rate was calculated for one set of background images. Each data point for Compton scatter and pair production sensitivity were generated by averaging the sensitivity over all simulated gamma-ray energies. All data here were generated using the full-scale simulation with a downsampling rate of 11.}  
\begin{tabular*}{0.9\columnwidth}{*3{C{0.25\columnwidth}}}
    \hline
     Background Rejection Rate (\%) & Pair Production Sensitivity (\%) & Compton Scatter Sensitivity (\%) \\
    \hline
    99.990$\pm$0.002 &  10$\pm$1  & 0.3$\pm$0.3 \\
    99.97$\pm$0.003   &  16$\pm$2 & 0.4$\pm$0.3 \\
    99.94$\pm$0.005   &  26$\pm$2 & 0.7$\pm$0.4 \\
    99.87$\pm$0.007   &  37$\pm$2 & 1.3$\pm$0.6 \\
    99.81$\pm$0.009   &  44$\pm$2 & 1.7$\pm$0.6 \\
    99.75$\pm$0.01   &  47$\pm$2 & 1.9$\pm$0.7 \\
    99.69$\pm$0.01   &  52$\pm$2 & 2.2$\pm$0.7 \\
    \hline
    \label{table:accuracies}
\end{tabular*}
\end{table}

\begin{figure}[bh!]
	\centering
	\captionsetup{justification=centering}	
	\begin{subfigure}[t]{0.48\textwidth}
		\centering
		\includegraphics[width=\textwidth]{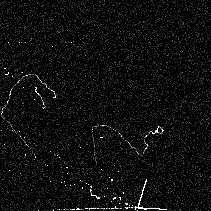}
		\caption{}
		\label{subfig:FP1}
	\end{subfigure}
	~
	\begin{subfigure}[t]{0.48\textwidth}
		\centering		
		\includegraphics[width=\textwidth]{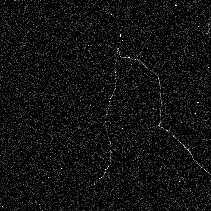}
		\caption{}
		\label{subfig:FP2}
	\end{subfigure}
	\\
	\begin{subfigure}[t]{0.48\textwidth}
		\centering
		\includegraphics[width=\textwidth]{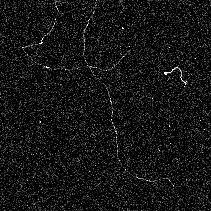}
		\caption{}
		\label{subfig:FP3}
	\end{subfigure}
	~
	\begin{subfigure}[t]{0.48\textwidth}
		\centering		
		\includegraphics[width=\textwidth]{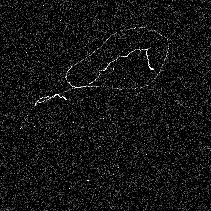}
		\caption{}
		\label{subfig:FP4}
	\end{subfigure}
	\caption{Projection images of simulated \ac{GCR} proton events that resulted in false positive classifications. Only the projection image resulting in the false positive is shown, the alternate projection is not included because no event produced a false positive in both projections.}
	\label{fig:FalsePositive}
\end{figure}

\begin{figure}[bh!]
	\centering
	\captionsetup{justification=centering}	
	\begin{subfigure}[t]{0.48\textwidth}
		\centering
		\includegraphics[width=\textwidth]{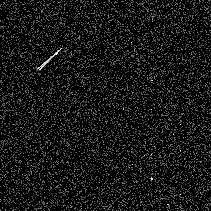}
		\caption{}
		\label{subfig:FN1}
	\end{subfigure}
	~
	\begin{subfigure}[t]{0.48\textwidth}
		\centering		
		\includegraphics[width=\textwidth]{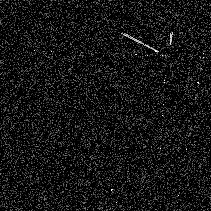}
		\caption{}
		\label{subfig:FN2}
	\end{subfigure}
	\\
	\begin{subfigure}[t]{0.48\textwidth}
		\centering
		\includegraphics[width=\textwidth]{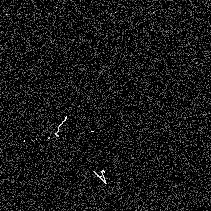}
		\caption{}
		\label{subfig:FN3}
	\end{subfigure}
	~
	\begin{subfigure}[t]{0.48\textwidth}
		\centering		
		\includegraphics[width=\textwidth]{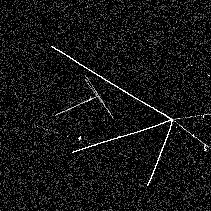}
		\caption{}
		\label{subfig:FN4}
	\end{subfigure}
	\caption{Projection images of pair production events that produced the lowest response in \texttt{GammaNet} for the pair production event class. The projection shown is most representative of the cause for false negative classification.}
	\label{fig:FalseNegative}
\end{figure}

In this study, the test set of \ac{GCR} protons contained 10$^6$ events, with twice as many images. Operating at 99.99\% background rejection resulted in 100 \ac{GCR} proton events being classified as positive, considered false positives events. False positives occur when at least one of two projections is classified as positive by \texttt{GammaNet}. \cref{fig:FalsePositive} shows 4 of the \ac{GCR} proton events that resulted in false positive classifications. \cref{fig:FalseNegative} shows 4 pair production events that resulted in \texttt{GammaNet} producing the lowest response for pair production classification out of the testing set. The projection shown for the \ac{GCR} proton events are the projection resulting in a positive classification, and the projections shown for the pair production events are the most representative of the characteristics resulting in false negative classification. In the false positive images, \cref{subfig:FP1,subfig:FP2,subfig:FP3,subfig:FP4}, extended delta-ray tracks are observed with at least one point of track crossing. This observation demonstrates that \texttt{GammaNet} responds to extended contiguous tracks, and track crossings, as signals of pair production events. In addition, \cref{subfig:FP1} contains a pair production event occurring from a \ac{GCR} proton track which results in a false positive classification, showing \texttt{GammaNet} responds significantly to the vertex of a pair production event. In the false negative images, \cref{subfig:FN1,subfig:FN2,subfig:FN3,subfig:FN4}, three features can be observed in the pair production images: short track length in \cref{subfig:FN1,subfig:FN2,subfig:FN3}, overlapping of the two tracks making it appear as a singular track in \cref{subfig:FN1,subfig:FN2}, and deep inelastic scattering events in \cref{subfig:FN4}.

		
\begin{figure}[th!]
    \centering
    \captionsetup{justification=centering}		
	\begin{subfigure}[t]{0.3\textwidth}
		\centering
		\includegraphics[width=\textwidth]{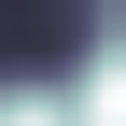}
		\caption{}
		\label{subfig:GCRCAM}
	\end{subfigure}
	~	
	\begin{subfigure}[t]{0.3\textwidth}
		\centering
		\includegraphics[width=\textwidth]{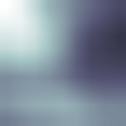}
		\caption{}
		\label{subfig:PPSCAM}
	\end{subfigure}		
	~
	\begin{subfigure}[t]{0.3\textwidth}
		\centering	
		\includegraphics[width=\textwidth]{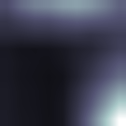}
		\caption{}
		\label{subfig:PPBCAM}
	\end{subfigure}	
	~~
	\begin{subfigure}[t]{0.3\textwidth}
		\centering
		\includegraphics[width=\textwidth]{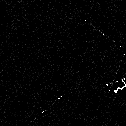}
		\caption{}
		\label{subfig:GCR}
	\end{subfigure}
	~	
	\begin{subfigure}[t]{0.3\textwidth}
		\centering
		\includegraphics[width=\textwidth]{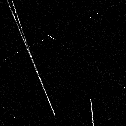}
		\caption{}
		\label{subfig:PPS}
	\end{subfigure}		
	~
	\begin{subfigure}[t]{0.3\textwidth}
		\centering	
		\includegraphics[width=\textwidth]{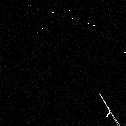}
		\caption{}
		\label{subfig:PPB}
	\end{subfigure}
	\caption{a-c) Images generated by the \ac{GCAM} algorithm that demonstrate the features that GammaNet utilizes for classifying images as background or signal. d-f) The simulation images used to generate the respective \ac{GCAM} images, with d) being a background event and e) and f) being signal events.}
	\label{fig:GCAM}
\end{figure}		
			
\vspace{-5mm}

\section{GammaNet Visualization}
\label{section:GradCAM}

As the use of \acp{CNN} becomes more prevalent in research, it is of increasing interest how the \ac{CNN} performs the classification and what features of the input it uses to do so. \ac{GCAM} \citep{GradCAM} is a recent algorithm developed to answer these questions by providing an activation map for input images that shows the regions the \ac{CNN} used most within the image during classification. \cref{fig:GCAM} shows the \ac{GCAM} images generated for GammaNet with one background image, \cref{subfig:GCR}, and two signal images, \cref{subfig:PPS,subfig:PPB}. These images were generated using the sub-scale simulation of \ac{AdEPT} because the lower track density provides interpretable results. \cref{subfig:GCRCAM} shows that for the background class, GammaNet utilized sparsely ionizing tracks and delta rays present in \cref{subfig:GCR}, resulting correctly in a background classification. \cref{subfig:PPSCAM} demonstrates that for the signal class, GammaNet utilizes the separate, nearly parallel, tracks of the pair production event preferentially over the overlapped pair production event at the bottom of \cref{subfig:PPS}, resulting in an accurate positive classification. Lastly, \cref{subfig:PPBCAM} results in a false background classification of the pair production image, \cref{subfig:PPB}, with GammaNet using the sparsely ionized \ac{GCR} tracks and the delta generated from the pair production track.

\section{Conclusion}

The event classification requirements of the \ac{AdEPT} mission dictate a background rejection rate between 99.99\% and 99.69\% which must be achieved within a 50 ms time window determined by the instrument collection rate. \texttt{GammaNet}, using mixed precision enabled by \texttt{NVCaffe}, was able to achieve a background rejection rate of 99.990$\pm$0.004\%. These results were achieved using the full-scale simulation, classifying on images downsampled at 11x. The time for inference was found to be on average 6.8 ms utilizing a NVIDIA GTX 1080 \ac{GPU}, which has 8.2 TFLOPS of single precision compute performance. This implies that, as is, \texttt{GammaNet} would require 1.1 TFLOPS of single precision compute available to it from the on-board flight computer. The \ac{AdEPT} mission is still in the development stages, and thus the flight computer has not been chosen. Commercially available flight computers are capable of meeting this demand. Additionally \ac{NASA} is investigating the use of commercial \ac{SoC} solutions that possess greater than 1 TFLOPS performance \cite{powell2018commercial}. In its current iteration, \texttt{GammaNet} is not prohibitively compute intensive for use as an on-board event classifier.

It was found that, in the best case situation of 99.69\% background rejection, signal sensitivity for pair production will range from 1.1$\pm$0.5\% to 69$\pm$2\% for 5 to 250 MeV incident gamma rays. This range becomes 0.1$\pm$0.1\% to 17$\pm$2\% for the worst case scenario requiring 99.99\% background rejection. The low sensitivity lowers the effectiveness of the \ac{AdEPT} instrument, however these values were generated using conservative estimates. These results show that \texttt{GammaNet} achieves the desired background rejections of \ac{AdEPT}, making it a serious consideration for use on-board the satellite for event classification.

These performance estimates include no image compression, and downlink bandwidth afforded by current and near future satellite communication \cite{atwood2009large,meegan2009fermi,cameron2012fermi,robinson2018terabyte}. No image compression was used as a conservative assumption due to the data handling system for the \ac{AdEPT} satellite not yet being decided. Simple lossless compression afforded by the PNG format produces compression ratios nearing 2 for the simulation images used in this study. As more systems aboard the \ac{AdEPT} satellite are designed and implemented, more precise determination of the operational parameters of \texttt{GammaNet} can be achieved. Reductions in the raw data rate will allow \texttt{GammaNet} to operate at a lower background rejection rate, affording increased pair production sensitivity. 

\ac{GCAM} was implemented for \texttt{GammaNet} in order to discern the features that \texttt{GammaNet} uses during classification of the simulation images. The results from this application support the supposition that \texttt{GammaNet} utilizes features that are characteristic of background for the respective classification, such as lower ionization density relative to the pair production tracks and the presence of delta rays. For the positive class of pair production the network responds strongly to semi-parallel tracks that are close in proximity, indicative of energetic pair production events.

\section{Appendix}
\subsection{\texttt{GammaNet} Architecture}
\label{subsection:appendix-architecture}

\begin{figure*}[h!]		
	\floatbox[{\capbeside\thisfloatsetup{capbesideposition={right,top},capbesidewidth=10cm}}]{figure}[\FBwidth]
	{
	\caption{Diagram depicting the architecture and layers used for \texttt{GammaNet}. All functions depicted in this diagram are from the preprogrammed operations included in the \texttt{NVCaffe} library. \\
	\\
	a) input to the network of an \ac{AdEPT} simulation image. \\
	\\
	b) first convolution layer made of a 7x7 convolution with a stride of 2, where stride is the spacing between the center of successive convolutions performed on the previous layer. The convolution is followed by a \ac{ReLU} operation, where all negative values are made to be 0. \\
	\\
	c) 3x3 max pooling layer with a stride of 2, where max pooling takes a subset of the previous layer and outputs the maximum value from that subset. The 3x3 max pooling is followed by a \ac{LRN} operation, where the values of the max pooling output are normalized along the depth of the output.\\
	\\
	d) 1x1 convolution with stride of 1 followed by a \ac{ReLU}. \\
	\\
	e) 3x3 convolution with a stride of 1 followed by a \ac{ReLU} and \ac{LRN} operation. \\
	\\
	f) 3x3 max pooling layer with a stride of 2. \\
	\\
	g) inception module used by \texttt{GoogLeNet} \cite{szegedy2015going}, part 1, from top to bottom is: 3 1x1 convolutions of stride 1 and a 3x3 max pooling with a stride of 1. \\
	\\
	h) inception module used by \texttt{GoogLeNet} \cite{szegedy2015going}, part 2, from top to bottom is a 3x3 convolution of stride 1, a 5x5 convolution of stride 1, and a 1x1 convolution of stride 1. \\
	\\
	i) concatenation along the depth of the previous 3 operations in h), where the separate outputs are combined into one 3 dimensional matrix. \\
	\\
	j) 7x7 average pooling with a stride of 1, where average pooling takes a subset of the previous layer and provides the average value for an output. The average pooling is followed by a dropout operation, where randomly some values in the output are set to 0 with a programmed probability. \\
	\\
	k) the flattening of j) into a vector. \\
	\\
	l) inner product between vector k) and the parameters of l) where there is a set of parameters for each class contained in the output, with two classes in the case of \texttt{GammaNet}. The parameters of l) were stored in double precision and the inner product calculated using double precision. \\
	\\
	m) 2 values output by the softmax operation, which takes the output of the inner product layer as an input for the softmax function. The softmax function provides the probability that the original input image belongs to each class of the network, pair production or background for \texttt{GammaNet}. The softmax function was calculated using double precision and its results were also produced with double precision.}\label{fig:NetworkDiagram}}
	{\includegraphics[width=7cm]{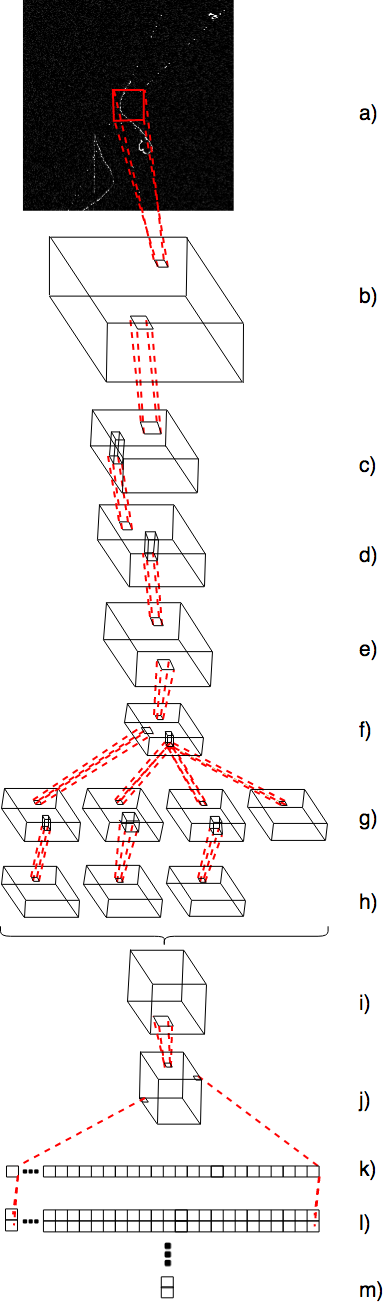}}
\end{figure*}

The architecture used for \texttt{GammaNet} is shown schematically in \cref{fig:NetworkDiagram} and is comprised of convolution, \ac{ReLU} \cite{nair2010rectified}, maximum or average pooling, \ac{LRN} \citep{krizhevsky2012imagenet,hinton2012improving}, dropout \cite{srivastava2014dropout}, concatenation, inner product, and softmax \citep{lawrence1997face,krizhevsky2012imagenet} operations. All of these operations come preprogrammed in \texttt{NVCaffe}, a platform for developing and programming \acp{CNN} \cite{jia2014caffe}, which was used for the development of \texttt{GammaNet}. 

\subsection{\texttt{GammaNet} Hyperparameters}
\label{subsection:Appendix-hyperparams}

The hyperparameters used in training \texttt{GammaNet} are as follows:

\noindent test\_iter: 1000 \\
test\_interval: 50000 \\
base\_lr: 0.0001 \\
display: 1000 \\
max\_iter: 10000000 \\
lr\_policy: "step" \\
gamma: 0.96 \\
momentum: 0.9 \\
weight\_decay: 0.0002 \\
stepsize: 320000 \\
snapshot: 49000 \\
snapshot\_prefix: "/path/to/your\_prefered\_directory" \\
solver\_mode: GPU \\
net: "/path/to/your\_network.prototxt" \\
test\_initialization: false \\
average\_loss: 40 \\
iter\_size: 1 \\

\section*{Acknowledgements}
This work has received support from the National Sciences and Engineering Research Council of Canada through research grants to S.H.B. Support for A.R.H was given by an appointment to the NASA Postdoctoral Program at the Goddard Space Flight Center administered by Oak Ridge Associated Universities through a contract with NASA. We gratefully acknowledge the support of NVIDIA Corporation with the donation of a \ac{GPU} which was used to accelerate computations for \texttt{GammaNet} in this research.

\bibliographystyle{elsarticle-num-names}
\bibliography{bibliography}

\begin{thebibliography}{10}
\expandafter\ifx\csname url\endcsname\relax
  \def\url#1{\texttt{#1}}\fi
\expandafter\ifx\csname urlprefix\endcsname\relax\def\urlprefix{URL }\fi

\bibitem[{Hoo-Chang et~al.(2016)Hoo-Chang, Roth, Gao, Lu, Xu, Nogues, Yao,
  Mollura, and Summers}]{hoo2016deep}
S.~Hoo-Chang, H.~R. Roth, M.~Gao, L.~Lu, Z.~Xu, I.~Nogues, J.~Yao, D.~Mollura,
  R.~M. Summers, {Deep convolutional neural networks for computer-aided
  detection: CNN architectures, dataset characteristics and transfer learning},
  IEEE transactions on medical imaging 35~(5) (2016) 1285.

\bibitem[{Szegedy et~al.(2017)Szegedy, Ioffe, Vanhoucke, and
  Alemi}]{szegedy2016inception}
C.~Szegedy, S.~Ioffe, V.~Vanhoucke, A.~A. Alemi, Inception-v4, inception-resnet
  and the impact of residual connections on learning, in: Thirty-first AAAI
  conference on artificial intelligence, 2017.

\bibitem[{Iandola et~al.(2016)Iandola, Han, Moskewicz, Ashraf, Dally, and
  Keutzer}]{iandola2016squeezenet}
F.~N. Iandola, S.~Han, M.~W. Moskewicz, K.~Ashraf, W.~J. Dally, K.~Keutzer,
  {Squeezenet: Alexnet-level accuracy with 50x fewer parameters and 0.5 mb
  model size}, arXiv preprint arXiv:1602.07360.

\bibitem[{Touvron et~al.(2020)Touvron, Vedaldi, Douze, and
  J{\'e}gou}]{touvron2020fixing}
H.~Touvron, A.~Vedaldi, M.~Douze, H.~J{\'e}gou, Fixing the train-test
  resolution discrepancy: Fixefficientnet, arXiv preprint arXiv:2003.08237.

\bibitem[{Russakovsky et~al.(2015)Russakovsky, Deng, Su, Krause, Satheesh, Ma,
  Huang, Karpathy, Khosla, Bernstein, Berg, and Fei-Fei}]{ILSVRC15}
O.~Russakovsky, J.~Deng, H.~Su, J.~Krause, S.~Satheesh, S.~Ma, Z.~Huang,
  A.~Karpathy, A.~Khosla, M.~Bernstein, A.~C. Berg, L.~Fei-Fei, {ImageNet}
  large scale visual recognition challenge, International Journal of Computer
  Vision (IJCV) 115~(3) (2015) 211--252.

\bibitem[{Baldi et~al.(2014)Baldi, Sadowski, and Whiteson}]{baldi2014searching}
P.~Baldi, P.~Sadowski, D.~Whiteson, {Searching for exotic particles in
  high-energy physics with deep learning}, Nature communications 5 (2014) 4308.

\bibitem[{Baldi et~al.(2016)Baldi, Bauer, Eng, Sadowski, and
  Whiteson}]{baldi2016jet}
P.~Baldi, K.~Bauer, C.~Eng, P.~Sadowski, D.~Whiteson, {Jet substructure
  classification in high-energy physics with deep neural networks}, Physical
  Review D 93~(9) (2016) 094034.

\bibitem[{Aurisano et~al.(2016)Aurisano, Radovic, Rocco, Himmel, Messier,
  Niner, Pawloski, Psihas, Sousa, and Vahle}]{NeutrinoCNN}
A.~Aurisano, A.~Radovic, D.~Rocco, A.~Himmel, M.~Messier, E.~Niner,
  G.~Pawloski, F.~Psihas, A.~Sousa, P.~Vahle, A convolutional neural network
  neutrino event classifier, Journal of Instrumentation 11~(09) (2016) P09001.

\bibitem[{Hunter et~al.(2014)Hunter, Bloser, Depaola, Dion, DeNolfo, Hanu,
  Iparraguirre, Legere, Longo, McConnell, et~al.}]{hunter2014pair}
S.~D. Hunter, P.~F. Bloser, G.~O. Depaola, M.~P. Dion, G.~A. DeNolfo, A.~Hanu,
  M.~Iparraguirre, J.~Legere, F.~Longo, M.~L. McConnell, et~al., {A pair
  production telescope for medium-energy gamma-ray polarimetry}, Astroparticle
  Physics 59 (2014) 18--28.

\bibitem[{Tavani et~al.(2009)Tavani, Barbiellini, Argan, Boffelli, Bulgarelli,
  Caraveo, Cattaneo, Chen, Cocco, Costa, et~al.}]{tavani2009agile}
M.~Tavani, G.~Barbiellini, A.~Argan, F.~Boffelli, A.~Bulgarelli, P.~Caraveo,
  P.~Cattaneo, A.~Chen, V.~Cocco, E.~Costa, et~al., {The AGILE mission},
  Astronomy \& Astrophysics 502~(3) (2009) 995--1013.

\bibitem[{Atwood et~al.(2009)Atwood, Abdo, Ackermann, Althouse, Anderson,
  Axelsson, Baldini, Ballet, Band, Barbiellini, et~al.}]{atwood2009large}
W.~Atwood, A.~A. Abdo, M.~Ackermann, W.~Althouse, B.~Anderson, M.~Axelsson,
  L.~Baldini, J.~Ballet, D.~Band, G.~Barbiellini, et~al., The large area
  telescope on the {Fermi} gamma-ray space telescope mission, The Astrophysical
  Journal 697~(2) (2009) 1071.

\bibitem[{Lei et~al.(1997)Lei, Dean, and Hills}]{lei1997compton}
F.~Lei, A.~Dean, G.~Hills, Compton polarimetry in gamma-ray astronomy, Space
  Science Reviews 82~(3-4) (1997) 309--388.

\bibitem[{Forot et~al.(2008)Forot, Laurent, Grenier, Gouiff{\`e}s, and
  Lebrun}]{forot2008polarization}
M.~Forot, P.~Laurent, I.~Grenier, C.~Gouiff{\`e}s, F.~Lebrun, Polarization of
  the crab pulsar and nebula as observed by the integral/ibis telescope, The
  Astrophysical Journal Letters 688~(1) (2008) L29.

\bibitem[{Bernard et~al.(2014)Bernard, Bruel, Frotin, Geerebaert, Giebels,
  Gros, Horan, Louzir, Poilleux, Semeniouk, et~al.}]{bernard2014}
D.~Bernard, P.~Bruel, M.~Frotin, Y.~Geerebaert, B.~Giebels, P.~Gros, D.~Horan,
  M.~Louzir, P.~Poilleux, I.~Semeniouk, et~al., {HARPO: a TPC as a gamma-ray
  telescope and polarimeter}, in: SPIE Astronomical Telescopes+
  Instrumentation, International Society for Optics and Photonics, 2014, pp.
  91441M--91441M.

\bibitem[{Takada et~al.(2011)Takada, Kubo, Nishimura, Ueno, Hattori, Kabuki,
  Kurosawa, Miuchi, Mizuta, Nagayoshi, et~al.}]{takada2011observation}
A.~Takada, H.~Kubo, H.~Nishimura, K.~Ueno, K.~Hattori, S.~Kabuki, S.~Kurosawa,
  K.~Miuchi, E.~Mizuta, T.~Nagayoshi, et~al., {Observation of diffuse cosmic
  and atmospheric gamma rays at balloon altitudes with an electron-tracking
  Compton camera}, The Astrophysical Journal 733~(1) (2011) 13.

\bibitem[{Ueno et~al.(2012)Ueno, Mizumoto, Hattori, Higashi, Iwaki, Kabuki,
  Kishimoto, Komura, Kubo, Kurosawa, et~al.}]{ueno2012development}
K.~Ueno, T.~Mizumoto, K.~Hattori, N.~Higashi, S.~Iwaki, S.~Kabuki,
  Y.~Kishimoto, S.~Komura, H.~Kubo, S.~Kurosawa, et~al., {Development of the
  balloon-borne sub-MeV gamma-ray Compton camera using an electron-tracking
  gaseous TPC and a scintillation camera}, Journal of Instrumentation 7~(01)
  (2012) C01088.

\bibitem[{Martoff et~al.(2005)Martoff, Ayad, Katz-Hyman, Bonvicini, and
  Schreiner}]{martoff2005negative}
C.~Martoff, R.~Ayad, M.~Katz-Hyman, G.~Bonvicini, A.~Schreiner, Negative ion
  drift and diffusion in a {TPC} near 1 bar, Nuclear Instruments and Methods in
  Physics Research Section A: Accelerators, Spectrometers, Detectors and
  Associated Equipment 555~(1-2) (2005) 55--58.

\bibitem[{Meegan et~al.(2009)Meegan, Lichti, Bhat, Bissaldi, Briggs,
  Connaughton, Diehl, Fishman, Greiner, Hoover, et~al.}]{meegan2009fermi}
C.~Meegan, G.~Lichti, P.~Bhat, E.~Bissaldi, M.~S. Briggs, V.~Connaughton,
  R.~Diehl, G.~Fishman, J.~Greiner, A.~S. Hoover, et~al., {The Fermi gamma-ray
  burst monitor}, The Astrophysical Journal 702~(1) (2009) 791.

\bibitem[{Cameron(2012)}]{cameron2012fermi}
R.~A. Cameron, Fermi large area telescope operations: progress over 4 years,
  in: SPIE Astronomical Telescopes+ Instrumentation, International Society for
  Optics and Photonics, 2012, pp. 84481J--84481J.

\bibitem[{Robinson et~al.(2018)Robinson, Boroson, Schieler, Khatri, Guldner,
  Constantine, Shih, Burnside, Bilyeu, Hakimi, et~al.}]{robinson2018terabyte}
B.~Robinson, D.~Boroson, C.~Schieler, F.~Khatri, O.~Guldner, S.~Constantine,
  T.~Shih, J.~Burnside, B.~Bilyeu, F.~Hakimi, et~al., Tera{B}yte {I}nfra{R}ed
  {D}elivery ({TBIRD}): a demonstration of large-volume direct-to-{E}arth data
  transfer from low-{E}arth orbit, in: Free-Space Laser Communication and
  Atmospheric Propagation XXX, Vol. 10524, International Society for Optics and
  Photonics, 2018, p. 105240V.

\bibitem[{Powell et~al.(2018)Powell, Campola, Sheets, Davidson, and
  Welsh}]{powell2018commercial}
W.~Powell, M.~Campola, T.~Sheets, A.~Davidson, S.~Welsh, Commercial
  off-the-shelf gpu qualification for space applications.

\bibitem[{Fawcett(2006)}]{fawcett2006introduction}
T.~Fawcett, An introduction to {ROC} analysis, Pattern recognition letters
  27~(8) (2006) 861--874.

\bibitem[{Agostinelli et~al.(2003)Agostinelli, Allison, Amako, Apostolakis,
  Araujo, Arce, Asai, Axen, Banerjee, Barrand, et~al.}]{agostinelli2003geant4}
S.~Agostinelli, J.~Allison, K.~a. Amako, J.~Apostolakis, H.~Araujo, P.~Arce,
  M.~Asai, D.~Axen, S.~Banerjee, G.~. Barrand, et~al., {GEANT4 a simulation
  toolkit}, Nuclear instruments and methods in physics research section A:
  Accelerators, Spectrometers, Detectors and Associated Equipment 506~(3)
  (2003) 250--303.

\bibitem[{Allison et~al.(2006)Allison, Amako, Apostolakis, Araujo, Dubois,
  Asai, Barrand, Capra, Chauvie, Chytracek, et~al.}]{allison2006geant4}
J.~Allison, K.~Amako, J.~Apostolakis, H.~Araujo, P.~A. Dubois, M.~Asai,
  G.~Barrand, R.~Capra, S.~Chauvie, R.~Chytracek, et~al., {Geant4 developments
  and applications}, IEEE Transactions on nuclear science 53~(1) (2006)
  270--278.

\bibitem[{Hanu(2018)}]{HanuSim}
A.~R. Hanu, {G4AdEPTSim},
  \url{https://github.com/AndreiHanu/G4AdEPTSim/releases} (2018).

\bibitem[{Badhwar(1997)}]{badhwar1997radiation}
G.~D. Badhwar, The radiation environment in low-earth orbit, Radiation research
  148~(5s) (1997) S3--S10.

\bibitem[{Benton and Benton(2001)}]{benton2001space}
E.~R. Benton, E.~Benton, Space radiation dosimetry in low-earth orbit and
  beyond, Nuclear Instruments and Methods in Physics Research Section B: Beam
  Interactions with Materials and Atoms 184~(1-2) (2001) 255--294.

\bibitem[{Zhou et~al.(2006)Zhou, O’Sullivan, Semones, and
  Heinrich}]{zhou2006radiation}
D.~Zhou, D.~O’Sullivan, E.~Semones, W.~Heinrich, Radiation field of cosmic
  rays measured in low earth orbit by cr-39 detectors, Advances in Space
  Research 37~(9) (2006) 1764--1769.

\bibitem[{Weidenspointner et~al.(2000)Weidenspointner, Varendorff, Kappadath,
  Bennett, Bloemen, Diehl, Hermsen, Lichti, Ryan, and
  Sch{\"o}nfelder}]{weidenspointner2000cosmic}
G.~Weidenspointner, M.~Varendorff, S.~Kappadath, K.~Bennett, H.~Bloemen,
  R.~Diehl, W.~Hermsen, G.~Lichti, J.~Ryan, V.~Sch{\"o}nfelder, The cosmic
  diffuse gamma-ray background measured with comptel, in: AIP Conference
  Proceedings, Vol. 510, American Institute of Physics, 2000, pp. 467--470.

\bibitem[{Henry(1999)}]{henry1999diffuse}
R.~C. Henry, Diffuse background radiation, The Astrophysical Journal Letters
  516~(2) (1999) L49.

\bibitem[{Ayres et~al.(2005)Ayres, Collaboration, et~al.}]{ayres2005nova}
D.~Ayres, N.~Collaboration, et~al., {NOvA proposal to build a 30 kiloton
  off-axis detector to study neutrino oscillations in the Fermilab NuMI
  beamline}, arXiv preprint hep-ex/0503053.

\bibitem[{Szegedy et~al.(2015)Szegedy, Liu, Jia, Sermanet, Reed, Anguelov,
  Erhan, Vanhoucke, and Rabinovich}]{szegedy2015going}
C.~Szegedy, W.~Liu, Y.~Jia, P.~Sermanet, S.~Reed, D.~Anguelov, D.~Erhan,
  V.~Vanhoucke, A.~Rabinovich, {Going deeper with convolutions}, in:
  Proceedings of the IEEE Conference on Computer Vision and Pattern
  Recognition, 2015, pp. 1--9.

\bibitem[{Jia et~al.(2014)Jia, Shelhamer, Donahue, Karayev, Long, Girshick,
  Guadarrama, and Darrell}]{jia2014caffe}
Y.~Jia, E.~Shelhamer, J.~Donahue, S.~Karayev, J.~Long, R.~Girshick,
  S.~Guadarrama, T.~Darrell, Caffe: Convolutional architecture for fast feature
  embedding, in: Proceedings of the 22nd ACM international conference on
  Multimedia, 2014, pp. 675--678.

\bibitem[{Simonyan and Zisserman(2014)}]{simonyan2014very}
K.~Simonyan, A.~Zisserman, Very deep convolutional networks for large-scale
  image recognition, arXiv preprint arXiv:1409.1556.

\bibitem[{Selvaraju et~al.(2017)Selvaraju, Cogswell, Das, Vedantam, Parikh, and
  Batra}]{GradCAM}
R.~R. Selvaraju, M.~Cogswell, A.~Das, R.~Vedantam, D.~Parikh, D.~Batra,
  {Grad-CAM}: Visual explanations from deep networks via gradient-based
  localization, in: Proceedings of the IEEE International Conference on
  Computer Vision, 2017, pp. 618--626.

\bibitem[{Nair and Hinton(2010)}]{nair2010rectified}
V.~Nair, G.~E. Hinton, {Rectified linear units improve restricted boltzmann
  machines}, in: Proceedings of the 27th international conference on machine
  learning (ICML-10), 2010, pp. 807--814.

\bibitem[{Krizhevsky et~al.(2012)Krizhevsky, Sutskever, and
  Hinton}]{krizhevsky2012imagenet}
A.~Krizhevsky, I.~Sutskever, G.~E. Hinton, {Imagenet classification with deep
  convolutional neural networks}, in: Advances in neural information processing
  systems, 2012, pp. 1097--1105.

\bibitem[{Hinton et~al.(2012)Hinton, Srivastava, Krizhevsky, Sutskever, and
  Salakhutdinov}]{hinton2012improving}
G.~E. Hinton, N.~Srivastava, A.~Krizhevsky, I.~Sutskever, R.~R. Salakhutdinov,
  {Improving neural networks by preventing co-adaptation of feature detectors},
  arXiv preprint arXiv:1207.0580.

\bibitem[{Srivastava et~al.(2014)Srivastava, Hinton, Krizhevsky, Sutskever, and
  Salakhutdinov}]{srivastava2014dropout}
N.~Srivastava, G.~Hinton, A.~Krizhevsky, I.~Sutskever, R.~Salakhutdinov,
  {Dropout: A simple way to prevent neural networks from overfitting}, The
  Journal of Machine Learning Research 15~(1) (2014) 1929--1958.

\bibitem[{Lawrence et~al.(1997)Lawrence, Giles, Tsoi, and
  Back}]{lawrence1997face}
S.~Lawrence, C.~L. Giles, A.~C. Tsoi, A.~D. Back, Face recognition: A
  convolutional neural-network approach, IEEE transactions on neural networks
  8~(1) (1997) 98--113.

\end{thebibliography}

\end{document}